# Lattice thermal expansion and anisotropic displacements in α-sulfur from diffraction experiments and first-principles theory

Janine George,[1] Volker L. Deringer,[1,†] Ai Wang,[1] Paul Müller,[1] Ulli Englert,[1,*] and Richard Dronskowski[1,2,*]

[1] Institute of Inorganic Chemistry, RWTH Aachen University, Landoltweg.1, Aachen 52074, Germany. E-mail: ullrich.englert@ac.rwth-aachen.de; drons@HAL9000.ac.rwth-aachen.de

[2] Jülich-Aachen Research Alliance (JARA-HPC), RWTH Aachen University, Aachen 52056, Germany

[†] Present address: Engineering Laboratory, University of Cambridge, Trumpington Street, Cambridge CB2 1PZ, United Kingdom

**Abstract:** Thermal properties of solid-state materials are a fundamental topic of study with important practical implications. For example, anisotropic displacement parameters (ADPs) are routinely used in physics, chemistry, and crystallography to quantify the thermal motion of atoms in crystals. ADPs are commonly derived from diffraction experiments, but recent developments have also enabled their first-principles prediction using periodic density functional theory (DFT). Here, we combine experiments and dispersion-corrected DFT to quantify lattice thermal expansion and ADPs in crystalline α-sulfur ($S_8$), a prototypical elemental solid that is controlled by the interplay of covalent and van der Waals interactions. We begin by reporting on single-crystal and powder X-ray diffraction (XRD) measurements that provide new and improved reference data from 10 K up to room temperature. We then use several popular dispersion-corrected DFT methods to predict vibrational and thermal properties of α-sulfur, including the anisotropic lattice thermal expansion. Hereafter, ADPs are derived in the commonly used harmonic approximation (in the computed zero-Kelvin structure) and also in the quasi-harmonic approximation (QHA) which takes the predicted lattice thermal expansion into account. At the PBE+D3(BJ) level, the QHA leads to excellent agreement with experiments. Finally, more general implications of this study for realistic materials modeling at finite temperature are discussed.

# I. INTRODUCTION

Modeling solid-state materials based on first-principles theory has become a central field of research and is routinely performed nowadays, most often in the framework of density-functional theory (DFT).[1, 2] Realistic models of solids, thereby, must go beyond zero Kelvin. This is done by including thermal vibrations (phonons) of the crystal lattice, typically described within the harmonic or the quasi-harmonic approximation.[3-7] In more recent years, also higher-order expansions of the potential energy (so-called anharmonic terms) are considered for phonon calculations.[8-12] However, this has mostly been done for small systems with high symmetry because it is computationally very demanding.

The motion of atoms in a crystal is also of fundamental interest, and it is quantified by anisotropic displacement parameters (ADPs) in crystallographic studies. Throughout solid-state theory, recent work demonstrated how ADPs can be predicted using periodic first-principles calculations, with applications ranging from prototypical solids such as diamond[13] to metal carbides[14] and layered chalcogenides[15], and further onward to molecular crystals[16-19] and organometallic compounds[20]. In crystallography, ADPs are particularly important because one can easily identify problems (such as disorder or twinning) in structure determination. Moreover, one can also use them to model thermal motion, derive translational and librational frequencies,[21, 22] and even calculate thermodynamic state functions such as the heat capacity $C_V$[23] or vibrational entropy.[24, 25] The field is thriving, but many open questions still remain to be fully resolved.

Besides first-principles prediction of ADPs also that of thermal expansion has been in the focus of research,[26-29] especially negative thermal expansion. In particular, some recent studies were devoted to the first-principles description of thermal expansion of molecular crystals.[5, 30]

Molecular crystals require a careful computational treatment including dispersion interactions, e.g. by dispersion-corrected DFT methods.[31, 32] To calculate ADPs, we and others[13, 16, 18-20] used periodic lattice dynamics in the harmonic approximation: the crystal structure is relaxed into the electronic ground state using an appropriate dispersion-corrected DFT method, phonon calculations are performed at the same level of theory, from which ADPs are then derived. Temperature hence enters through the statistics used to determine ADPs, but the influence of temperature on the lattice expansion is neglected. Considering lattice expansion changes the phonon frequencies (see, e.g., a detailed study on lead chalcogenides[10]) and thus will change the calculated ADPs. In the past, we have considered the volume expansion by simply using the experimental lattice parameters,[18] but this can only be an ad-hoc improvement and only in cases where the volume expansion is faithfully described by DFT. Madsen et al.[16] took this effect into account by using experimentally derived and fitted Grüneisen parameters to scale frequencies within a quasi-harmonic framework.

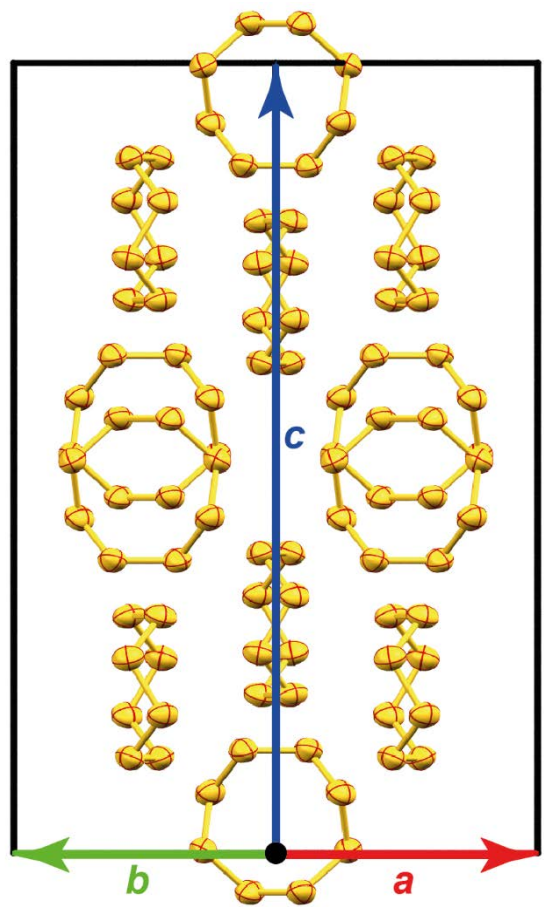

FIG. 1. Conventional face-centered ($Z = 16$) unit cell of α-sulfur;[33] the single-crystal diffraction result at 100 K is shown. The conventional view direction [110] shows both side views and top views of the puckered $S_8$ rings. Displacement ellipsoids are drawn at the 99% probability level.

In this work, we study crystalline, elemental sulfur in its α form that is stable at ambient conditions (Fig. 1). This allotrope crystallizes in the orthorhombic space group *Fddd* with 16 molecules in a face-centered unit cell. The asymmetric unit contains four sulfur atoms in general position, close to a twofold axis which generates the entire puckered $S_8$ ring. The approximate symmetry of such a crown-shaped $S_8$ molecule is $D_{4d}$. Shortest contacts between neighboring cyclooctasulfur residues amount to 3.3 Å. The rather small asymmetric unit, the absence of strongly absorbing atom types and the dispersion-dominated intermolecular interactions make α-$S_8$ a very well-suited candidate for calculations with dispersion-corrected DFT and concomitant experimental evaluation. There have been several DFT studies on α-sulfur so far,[34, 35] including a recent application of the quasi-harmonic approximation (QHA) to calculate thermodynamic properties,[36] to which we will return below. In this work, we measure experimentally and calculate from first principles the lattice thermal expansion and ADPs of α-$S_8$. For the latter purpose, we go beyond the simple harmonic approximation and predict ADPs in the QHA, thereby including the first-principles derived thermal expansion of the crystal.

## II. THEORY

We calculate harmonic vibrational frequencies and normalized phonon eigenvectors using the finite displacement method as implemented in PHONOPY[37, 38], imposing individual atomic displacements of 0.01 Å. Vibrational frequencies and eigenvectors are then utilized to calculate the mean-square displacement matrices. This is also implemented in PHONOPY[18, 37, 38]. Next, the mean-square displacement matrices are transformed into anisotropic displacement tensors. Thereby, the coordinate system is changed from the Cartesian to the crystal coordinate system.[19, 39] In crystallography, these values $U_{ij}$ are used within the Debye-Waller factor DWF[40] that accounts for thermal movement of the individual atoms and the attenuation of X-ray and coherent neutron scattering:

$$\text{DWF} = e^{-2\pi^2\left(\sum_{i=1}^{3}\sum_{j=1}^{3} h_i a_j U_{ij} a_j h_j\right)} \quad (1)$$

where $h_i$ is the $i^{\text{th}}$ component of the diffraction vector and $a_i$ is the $i^{\text{th}}$ component of the reciprocal lattice vector.

Next, we calculate the main-axis components $U_1$, $U_2$, $U_3$ of the full anisotropic displacement tensor; finally, these are used to obtain the equivalent isotropic displacement parameter $U_{\text{eq}}$ (a quantity comparable to the isotropic displacement parameter $U_{\text{iso}}$):

$$U_{\text{eq}} = \frac{U_1 + U_2 + U_3}{3} \quad (2)$$

This is carried out by a custom-written program[19, 41] based on the formulas in Ref. 39.

To predict thermal expansion, we apply a slightly modified variant of the QHA[3] as detailed in Ref. 42 that was previously used successfully to calculate thermal properties of oxide materials[43-45] and molecular crystals[46]. In doing so, we start with the fully energetically optimized structure at the ground-state volume (denoted $V_0$). Then, we calculate an energy–volume curve $E_0(V)$ at volumes from $0.94^3 \times V_0$ to $1.06^3 \times V_0$ in steps of $0.01^3 \times V_0$, with a constant-volume optimization at each increment. The lattice parameters and atom positions are therefore optimized by minimizing the electronic energy.

The vibrational part of the Helmholtz free energy $F_{\text{vib}}$ is then calculated at volumes from $0.96^3 \times V_0$ to $1.00^3 \times V_0$ in steps of $0.01^3 \times V_0$ and at several temperatures $T$. Hereafter, $F_{\text{vib}}$ is linearly fitted at each temperature and these fits are then utilized to calculate $F_{\text{vib}}$ for the volumes from $0.94^3 \times V_0$ to $1.06^3 \times V_0$ in steps of $0.01^3 \times V_0$. This is the only difference to the traditional QHA: in the latter, $F_{\text{vib}}$ is calculated at each increment with the help of the harmonic approximation, and no linear fit is performed. The explained modification saves computing time compared to the traditional QHA; it can also lead to improvements in cases where the course of $F_{\text{vib}}(V)$ is not steady, e.g. due to the appearance of soft modes[42]. In return, however, one must carefully check for possible inaccuracies when using this modified QHA because only volumes smaller or equal to the equilibrium volume are considered and $F_{\text{vib}}$ is expected to depend approximately linearly on the volume. To assess the validity of the approximation made here, we also performed a traditional QHA computation (as implemented in PHONOPY)[37, 38] at that level of theory that reproduces the experimental values best (Supplementary Material): the influence of the present approximation on the predicted thermal expansion is very small (volume differences <0.1 Å³/atom). We also note in passing that there exist other strategies to save computing time within the QHA, by evaluating $F_{\text{vib}}$ at only a small number of both compressed and expanded volume increments.[6, 47]

Once the vibrational contributions are known, the Helmholtz free energy $F(V;T)$ is calculated according to

$$F(V;T) = E_0(V) + F_{\text{vib}}(V;T) \tag{3}$$

and subsequently, fitting $F(V;T)$ to the Murnaghan equation of state at each temperature gives the optimal volume at the different temperatures:

$$V_0(T) = \underset{V}{\arg\min}(F(V;T)). \tag{4}$$

To determine lattice parameters corresponding to each of the volumes, we use VASP to relax the structures in order to minimize energy under the constraint of constant volumes $V_0(T)$:

$$(a_0(T), b_0(T), c_0(T)) = \underset{(a,b,c)}{\arg\min}(E(a,b,c)), \text{subject to } a \cdot b \cdot c = V_0(T) \tag{5}$$

For structural optimizations of the conventional cell, we employed 4×4×2 $k$-points, whereas for the phonon calculations (QHA and ADPs) we used 2×2×1 supercells of the conventional unit cell (Fig. 1) with reciprocal-space sampling at Γ. The reciprocal mesh (*q*-mesh) for the calculation of the vibrational part of the free energy $F_{vib}$ is 48×48×24; for the ADP calculation, it is increased to 52×52×26. In the ADP calculation, we also cut off frequencies lower than 0.1 THz, as described in previous work[19].

Both for structural optimizations and phonon calculations, the forces on atoms were obtained using dispersion-corrected DFT as implemented in VASP[48-51], with strict convergence criteria of $\Delta E < 10^{-7}$ ($10^{-5}$) eV per cell for electronic (structural) optimizations, respectively. Moreover, we use the projector augmented-wave[52, 53] method with a plane wave cutoff of 500 eV and the following functionals with dispersion-corrections: the Perdew–Burke–Ernzerhof (PBE) functional augmented by the "D3" correction of Grimme and co-workers together with Becke–Johnson damping, denoted PBE+D3(BJ) in the following;[54-56] the scheme of Tkatchenko and Scheffler which builds on the same functional but uses system-dependent dispersion coefficients, denoted PBE+TS[54, 57, 58]; finally, we use the standalone van der Waals density functional vdW-DF2[59-62]. This selection is based on our previous studies[19, 20] where ADPs calculated by these particular techniques agreed well with experiment. By contrast, the uncorrected PBE functional is known to overestimate each lattice parameter of the conventional cell of orthorhombic sulfur by more than 6%,[34] and it is hence not used in the following.

### III. EXPERIMENTAL METHODS

The orthorhombic α allotrope of sulfur is stable below 95 °C;[63] the β phase melts at 115 °C.[63] Therefore, ADPs are only evaluated up to 200 K (−73 °C).

Different samples of α-$S_8$ were used for single crystal and powder X-ray diffraction. A light yellow rod with approximate dimensions 0.22×0.15×0.08 mm³ was chosen directly from a commercial batch (Merck GmbH, Darmstadt, Germany). This crystal was used for diffraction experiments at 100, 150 and 200 K; the temperature was controlled with an Oxford Cryostream 700 instrument. Intensity data were collected on a Bruker D8 goniometer with a SMART APEX CCD area detector in ω-scan mode using Mo-Kα radiation (λ = 0.71073 Å) from an Incoatec microsource with multilayer optics.

Reflections were integrated with SAINT+ [64] and multi-scan absorption corrections were applied with SADABS[65]. Before structure refinement, the lattice parameters from the single-crystal intensity integration process were adjusted to the more reliable values obtained from powder diffraction experiments at the same temperature. The atoms were assigned anisotropic displacement parameters, and refinements were accomplished with full-matrix least-square procedures based on $F^2$ as implemented in SHELXL-14[66]. The CIF files for the refinements at 100 (standard and high resolution), 150, and 200 K were deposited at the Cambridge Crystallographic Data Centre and allocated the following reference numbers: CCDC 1506778, 1506779, 1506777, and 1506776, respectively.

The α-$S_8$ batch from which the single crystal of very good quality was chosen proved challenging for powder diffraction due to very strong preferred orientation: no satisfactory displacement parameters could be obtained. Scanning electron microscopy (Fig. 2) confirmed that an alternative sample from desulfurization in a biorefinery (see Supplementary Material for more detailed information about this sample) contained beautiful microcrystals of tetragonal-bipyramidal morphology; this material was used for powder diffraction.

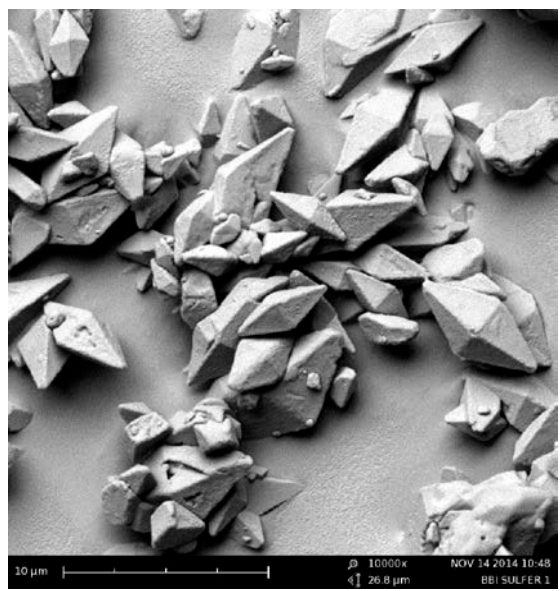

FIG. 2: Scanning electron microscopic image of the α-$S_8$ sample obtained from desulfurization in a biorefinery.

Flat powder samples were measured using a Guinier diffractometer (G645, Huber, Rimsting; CuKα$_1$ radiation, Ge Johansson-type monochromator) with a closed-cycle refrigerator in the temperature range 10 – 298 K, controlled via a Si diode. The scattered intensities were detected by a scintillation counter in the $2\theta$ range 8–100° in steps of 0.02° (20 s/step). Scans were measured at 298, 250, 200, 150, 100, 50, and 10 K, respectively; the profiles were fitted using the Rietveld method as implemented in the Fullprof Suite[67]. Figure 3 shows the measured and calculated intensities for the scan at 10 K. The lattice parameters and the isotropic displacement parameters resulting from the profile fitting are shown in the Supplementary Material.

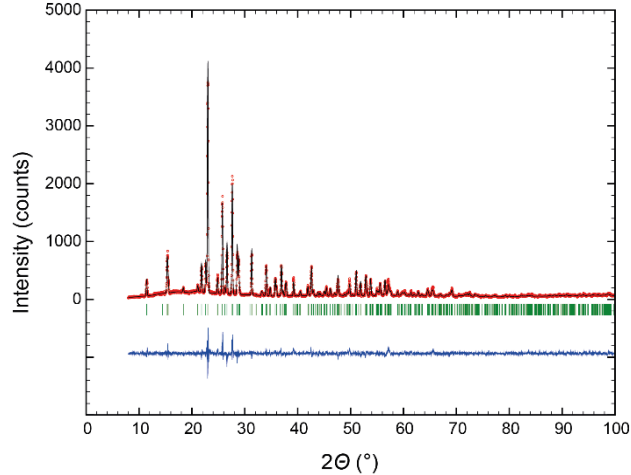

FIG. 3: Measured (red) and calculated (black) XRD intensities of α-S$_8$ powder at 10 K. The blue line below is a difference plot which confirms a highly reliable refinement result.

## IV. RESULTS AND DISCUSSION

### A. Ground-State Properties and Lattice Dynamics

TABLE 1. Unit-cell volumes, averaged intra-ring bond lengths $\bar{d}_{\text{S–S}}$, and averaged inter-ring bond lengths $\bar{d}_{\text{S}\cdots\text{S}}$ of crystalline α-sulfur. Results were obtained by optimization of the total energy at different levels of theory (not including ZPE corrections), and from experiment (powder X-ray diffraction at 10 K as detailed in Sec. III), as well as the root mean square (RMS) of the Cartesian deviations between calculated structures and the experimental one at 10 K. [68, 69] The labels of the sulfur atoms in the footnotes are consistent with Figure 9 and with the CIF files provided in the Supplementary Material.

| Method | $V$ (Å$^3$ atom$^{-1}$) | $\bar{d}_{\text{S–S}}$ (Å)$^a$ | $\bar{d}_{\text{S}\cdots\text{S}}$ (Å)$^b$ | RMS (Å) |
|---|---|---|---|---|
| vdW-DF2 (0 K) | 28.04 | 2.108 | 3.484 | 0.05 |
| PBE+TS (0 K) | 27.62 | 2.060 | 3.493 | 0.08 |
| PBE+D3(BJ) (0 K) | 24.73 | 2.058 | 3.294 | 0.04 |
| Expt. (10 K; this work) | 25.0214(16) | 2.058(5) | 3.333 | − |

$^a$ $\bar{d}_{\text{S−S}} = \frac{1}{5}(d(\text{S1–S1'}) + d(\text{S1–S2}) + d(\text{S2–S3}) + d(\text{S3–S4}) + d(\text{S4–S4'}))$.
$^b$ $\bar{d}_{\text{S}\cdots\text{S}} = \frac{1}{2}(d(\text{S1}\cdots\text{S1'}) + d(\text{S2}\cdots\text{S2'}))$.

We start with structural optimization, minimizing the total energy of α-S$_8$, and compare the computed volumes to our experimental results at 10 K (Table 1). Among the computational methods, vdW-DF2 and PBE+TS overestimate the experimental volume quite drastically (12% and 10%, respectively), even without considering the zero-point energy (ZPE). In contrast, the PBE+D3(BJ) volume fits well to the experimental volume, to within 1.5%. In a previous study on crystalline pentachloropyridine, we did not observe any correlation between the agreement of the volume reached at minimal energy and the ADP description,[19] but the differences in the equilibrium volume were not as drastic as they are here. The lengths of the S−S bonds are well described at the PBE+TS and PBE+D3(BJ) levels of theory, whereas vdW-DF2 leads to slightly too long S−S distances (Table 1). We also compare all secondary, *intermolecular* S···S distances smaller than the sum of the van der Waals radii in the experimental structure at 10 K. As already expected from the calculated volumes, PBE+D3(BJ) best reproduces the experimental distances (Table 1) whereas PBE+TS and vdW-DF2 overestimate them notably. As an additional quality criterion for the computationally established structure model, we use the root mean square (RMS) of the Cartesian deviations,[68,69] as defined in Ref. 68. PBE+D3(BJ) shows the lowest value, again suggesting best agreement with experiment.

TABLE 2. Frequencies of external modes computed in the harmonic approximation (this work) compared to data from vibrational spectroscopy at temperatures below 20 K (data from Ref. 70). The mean absolute deviation (MAD) is given below.

| Irreducible Representations | Expt. (cm$^{-1}$) (Ref. 70) | PBE+D3(BJ) (cm$^{-1}$) | PBE+TS (cm$^{-1}$) | vdW-DF2 (cm$^{-1}$) |
|---|---|---|---|---|
| $B_{2g}$ | 31.4 | 30.2 | 29.9 | 26.2 |
| $B_{3g}$ | 32.6 | 31.6 | 27.0 | 26.2 |
| $B_{2u}$ | 35.74 | 34.1 | 33.4 | 31.9 |
| $B_{3u}$ | 33.76 | 33.1 | 32.6 | 28.3 |
| $A_g$ | 54.4 | 52.8 | 45.8 | 47.0 |
| $B_{1g}$ | 40.8 | 38.3 | 40.6 | 33.3 |
| $B_{2g}$ | 54.4 | 53.4 | 46.5 | 47.8 |
| $B_{3g}$ | 45.6 | 43.2 | 47.1 | 39.3 |
| $A_u$ | − | 60.0 | 52.8 | 56.8 |
| $B_{1u}$ | 41.75 | 39.0 | 49.7 | 40.6 |
| $B_{2u}$ | 54.48 | 50.5 | 39.3 | 41.5 |
| $B_{3u}$ | 53.13 | 50.5 | 40.6 | 45.1 |
| MAD (cm$^{-1}$) |  | 1.9±1.0 | 5.8±5.0 | 6.4±2.9 |

The foundation of all analyses in this work is a description of the crystal's lattice dynamics, as expressed through the phonon eigenvalues. We hence begin by assessing the quality of the calculated frequencies at the different levels of theory. Therefore, we compare the lowest calculated, mostly external modes ($T = 0$ K, without ZPE) with experimental results from vibrational spectroscopy below 20 K (Table 2).[70] Unfortunately, we cannot compare all frequencies because some phonon modes accessible by IR and Raman are still unassigned.

Nearly all calculated vibrational frequencies are smaller than their experimental counterparts. This does not hold for all frequencies at the PBE+TS level of theory. Moreover, the mean absolute deviation for PBE+D3(BJ), (1.9±1.0) cm$^{-1}$, is significantly smaller than that at the other levels of theory. In conclusion, the PBE+D3(BJ) method performs best among the three investigated when it comes to predicting the external vibrational modes of crystalline α-sulfur.

To also assess internal frequencies, we additionally compare all phonon modes with $A_g$ symmetry (Table 3). Again, PBE+D3(BJ) fits best to experiment; PBE+TS follows closely, whereas vdW-DF2 deviates the most among the three methods under study. In this comparison, the ranking of the different levels of theory is even clearer than in the comparison before. Nearly all compared experimental frequencies are underestimated by the different levels of theory. The good performance of PBE+D3(BJ) is not surprising: a similar comparison of vibrational frequencies for benzene and other dispersion-dominated crystals has already shown PBE+D3(BJ)'s good agreement with experimental vibrational frequencies.[71]

TABLE 3. Comparison of the frequencies of modes with the irreducible representation $A_g$ from the harmonic approximation and from vibrational spectroscopy below 20 K (data from Ref. 70).

| Expt. (cm$^{-1}$)[70] | PBE+D3(BJ) (cm$^{-1}$) | PBE+TS (cm$^{-1}$) | vdW-DF2 (cm$^{-1}$) |
|---|---|---|---|
| 54.4 | 52.8 | 45.8 | 47.0 |
| 58.4 | 54.7 | 52.8 | 48.9 |
| 91.6 | 93.3 | 92.0 | 87.0 |
| 156.2 | 148.3 | 143.7 | 141.6 |
| 198 | 194.4 | 191.1 | 184.1 |
| 218.6 | 217.5 | 217.6 | 204.5 |
| 246.8 | 245.4 | 246.7 | 234.1 |
| 416.2 | 357.3 | 351.3 | 272.6 |
| 440.3 | 393.5 | 387.1 | 321.3 |
| 470 | 453.6 | 450.9 | 398.3 |
| 475 | 460.1 | 459.2 | 413.0 |
| 476.8 | 468.6 | 469.2 | 426.6 |
| MAD (cm$^{-1}$) | 13.9±19.1 | 15.7±21.4 | 39.4±48.7 |

We also compare the frequencies derived by Shang et al.[36] with their favored method in their previous study on sulfur, PBEsol+D3, in the same ways. The frequencies at this level of theory are approximately of the same quality as our PBE+D3(BJ) results; at the PBEsol+D3 level, the previous authors determined frequencies for which we here calculate MADs of 2.03±2.08 cm$^{-1}$ (≈ external phonon modes) and 11.21±14.06 cm$^{-1}$ (irreducible representation $A_g$). To save computing time and for better consistency with our previous studies, we will not use the latter method in this work. There is also another reason why we will not utilize it, to which we will come back later on.

## B. ADPs in the Harmonic Approximation

We now turn to the calculation of ADPs in the harmonic approximation—that is, with cell vectors frozen to their respective ground-state values. This method is commonly used in current work on ADP prediction.[13, 16, 18-20]

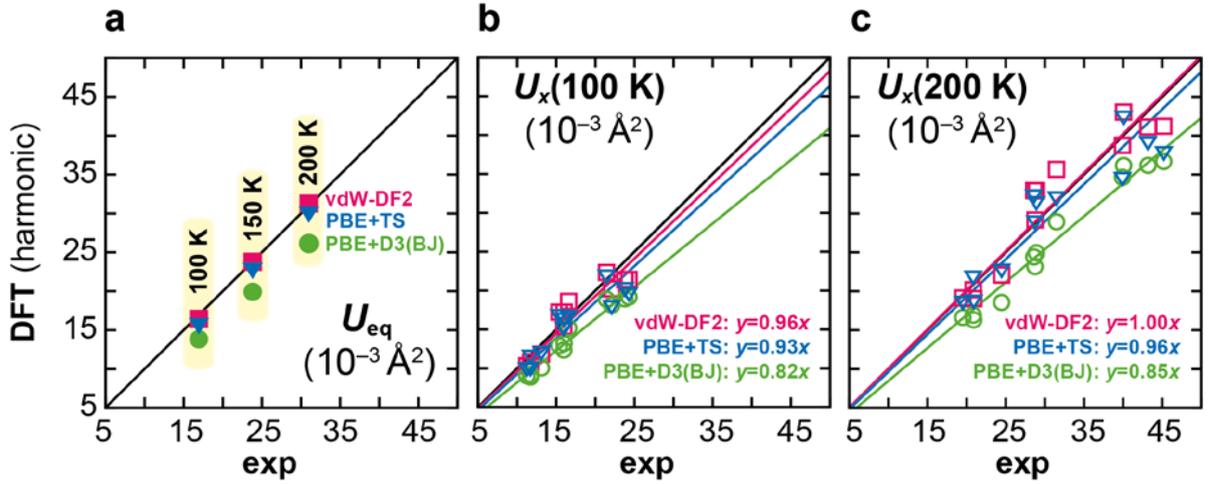

FIG. 4. (a) Equivalent displacement parameters ($U_{eq}$) of α-sulfur in the harmonic approximation. (b–c) Same for the main-axis components of the anisotropic displacement tensor at 100 K and 200 K. Results of a linear least-squares fit to the data points are noted; the diagonal identity is given by a solid black line here and in the following.

We have gathered experimental benchmark values at 100, 150, and 200 K using single-crystal XRD, and use these to assess our computed values at the different levels of theory (Fig. 4a). The equivalent displacement parameters $U_{eq}$ shown there are given as arithmetic means over the four crystallographically independent sulfur atoms. Surprisingly at first sight, the quality of predictions is reversed compared to the above frequency computations (Tables 2 and 3): in the ADP case, the vdW-DF2 functional seems to fit the experimental data perfectly; PBE+TS looks nearly equally well, whereas PBE+D3(BJ) shows the most significant deviations. Moreover, the $U_{eq}$ from vdW-DF2 are largest, then the ones from PBE+TS follow, and finally those from PBE+D3(BJ). The latter trend clearly correlates with what was seen for the unit-cell volumes at these levels of theory (Table 1).

To gain more detailed insight in the description of individual ADPs and their anisotropy, we also compare the calculated and experimental main-axis components for all sulfur atoms at 100 K (Fig. 4b) and 200 K (Fig. 4c), by linear regression ($y = m \times x$). For PBE+D3(BJ), the slope $m$ of this fit deviates the most from unity and, therefore, its agreement with experiment is worst, as already been expected from $U_{eq}$. Also in the comparison of the main-axis components, vdW-DF2 shows best agreement with experiment.

Overall, the underestimation of the frequencies $\omega_i$ at the vdW-DF2 level of theory could lead to vdW-DF2's accidental good performance for the calculation of ADPs at higher temperatures: the magnitude of the ADPs is proportional to $1/\omega_i$ and therefore if $\omega_i$ are underestimated, the harmonic ADPs will be overestimated.

To sum up to this point, PBE+D3(BJ) performs very well for vibrations at low temperature (Tables 2 and 3), but the agreement with experiment regarding ADPs at 100 K is less satisfactory (Fig. 4). Therefore, the question arises whether the quality of the calculated ADP at this level of theory could be improved by considering volume expansion.

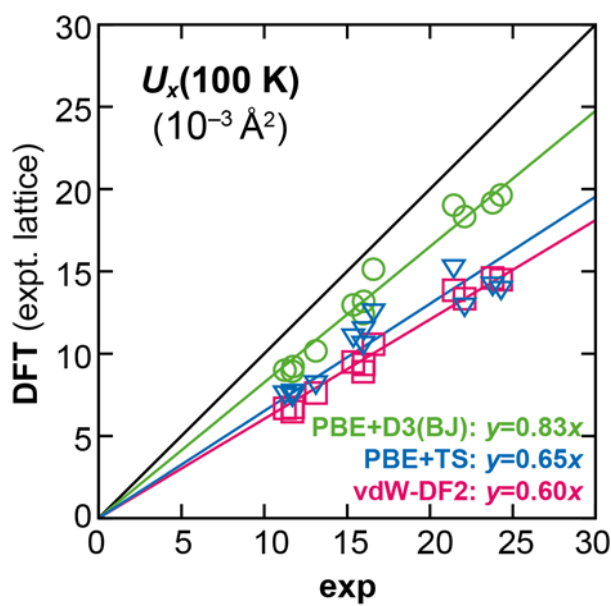

FIG. 5. Main-axis components of the anisotropic displacement tensor at 100 K of α-sulfur calculated at experimental lattice constants (see Supplementary Material). Results of a linear least-squares fit to the data points are noted.

To do so, we first attempt an ad-hoc improvement as discussed in earlier work[18]: we calculate ADPs at 100 K at the experimentally derived lattice constants (single-crystal data; see Supplementary Material for details). This volume is compressed by 12% and 10% relative to the vdW-DF2 and PBE+TS minima, respectively, while it nearly corresponds to the computed value for PBE+D3(BJ) (cf. Table 1). Therefore, calculated ADPs will agree well with experiment if, and only if, the DFT methods are well-suited to describe thermal expansion and energetics at such an expanded volume. Indeed, in this case PBE+D3(BJ) agrees best with experiment (Fig. 5); PBE+TS and vdW-DF2 perform significantly worse.

This again suggests that the good performance of vdW-DF2 in the purely harmonic approximation originates from error compensation. To gain more insight in the thermal expansion as described at these levels of theory and its influence on the ADP calculation, we apply the QHA.

### C. Lattice Expansion and Quasi-Harmonic Approximation

Figure 6 shows the unit-cell volumes as a function of temperature, as calculated at the different levels of theory in the simplified QHA. The PBE+D3(BJ) calculation is clearly in best agreement with experiment among the three. Both vdW-DF2 and PBE+TS, by contrast, overestimate the experimental volume massively, as had already been suggested by the zero-Kelvin structural optimization (Table 1). Moreover, all levels of theory overestimate the volume expansion, as seen from the calculated slope for the volume-temperature plots (Table 4). To compare the volumetric thermal expansion in more detail, we show the volumetric thermal expansion coefficients in the Supplementary Material; all levels of theory overestimate the experimental coefficients, more strongly so at higher temperature. PBEsol+D3, as used in Ref. [36], drastically underestimates the experimental volume—yielding 23.61 Å$^3$/atom when including zero-point energy, and thus a value 6% below our experimental result at 10 K (Table 1). Although this method displays good agreement with experiment regarding *relative* volume expansion,[36] the *absolute* volume is not reproduced correctly at this level of theory. Therefore, it does not seem well-suited for the present purpose of predicting the absolute volume expansion.

To analyze the thermal expansion for the most suitable level of theory, PBE+D3(BJ), in more detail, we also investigated the anisotropic thermal expansion (Figure 7). At lower temperatures, all calculated lattice parameters are in good agreement with their experimental counterparts (deviations for all lattice parameters < 0.5% at 50 K). The discrepancies increase at higher temperatures. At around 300 K, there is still satisfactory agreement between experiment and theory; the deviations are still lower than 1.2%. Furthermore, the relative deviation of the lattice parameter *a* is more pronounced than that of *b* and *c* at this temperature.

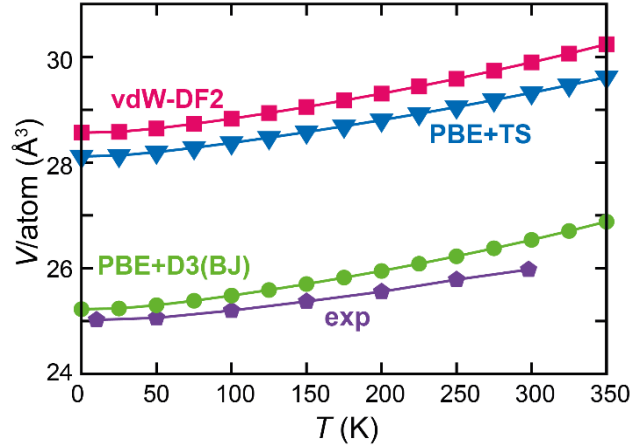

FIG. 6. Volume expansion of α-sulfur with increasing temperature as determined from powder XRD experiments and calculated within the QHA. The experimental error bars are smaller than the points used in this graph.

TABLE 4. Slope of volume-temperature plot $\frac{\Delta V}{\Delta T}$ of crystalline α-sulfur derived by linear regression of the volume in the temperature-range from 50 to 300 K.

| Method | $\frac{\Delta V}{\Delta T}$ (Å³/K) | |
| --- | --- | --- |
| Expt. | 0.0037 | ± 0.0002 |
| PBE+D3(BJ) | 0.0050 | ± 0.0001 |
| vdW-DF2 | 0.0050 | ± 0.0001 |
| PBE+TS | 0.0045 | ± 0.0001 |

Moreover, we calculate the slope of each lattice parameter-temperature plot and compare the results to experiment (Table 5). While the slope for the *b* parameter agrees well with experiment, that for *a* is predicted too large by 57 %, and that for *c* by 51 %. The deviations might be explainable by an increasing inaccuracy of PBE+D3(BJ) when atoms are moved away from their equilibrium position.[72] Also, anharmonic effects beyond the quasi-harmonic approximation are neglected and could possibly explain these deviations.[73, 74] It should be noted, however, that cases exist where the QHA works well even up to the melting temperature of a given compound;[75] clearly, these predictions are system-dependent. To conclude, we can also simulate the lattice expansion in qualitatively correct agreement with experiment. This is a necessary requirement for an accurate ADP calculation.

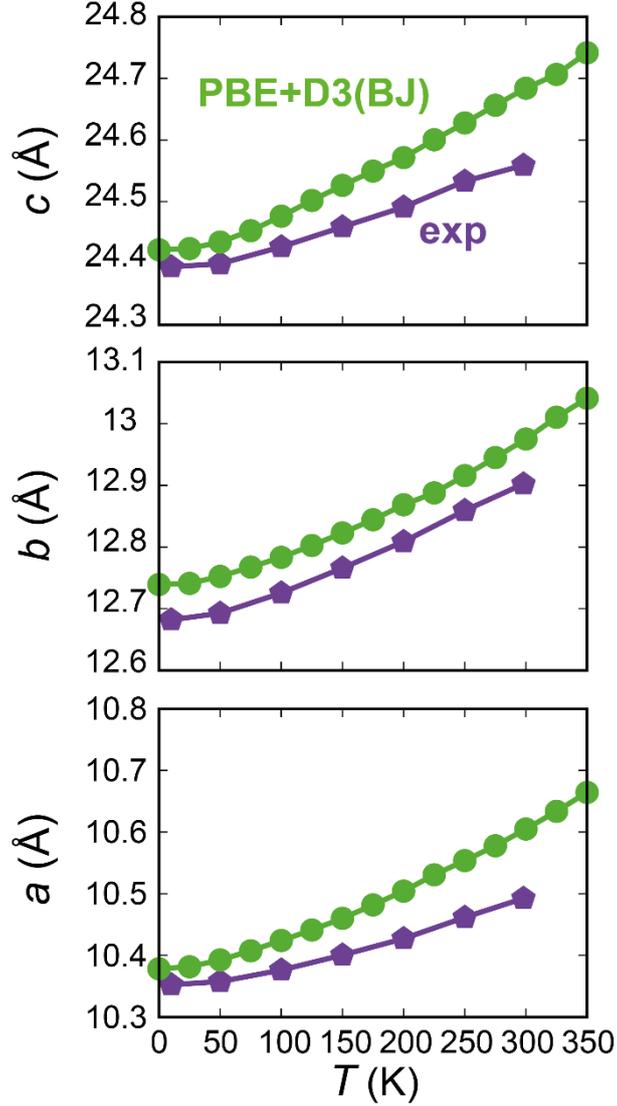

FIG. 7. Thermal expansion of the individual unit-cell parameters in α-sulfur. Our experimental data from powder XRD (Sec. III) are given for comparison.

TABLE 5. Slopes of lattice parameter vs temperature plots of crystalline α-sulfur, derived by linear regression of the lattice parameters $a$, $b$ and $c$ in the temperature-range from 50 to 300 K.

| Method | Expt. | PBE+D3(BJ) |
|---|---|---|
| $\frac{\Delta a}{\Delta T}$ (Å/K) | 0.00054± 0.00003 | 0.00085±0.00003 |
| $\frac{\Delta b}{\Delta T}$ (Å/K) | 0.00085± 0.00002 | 0.00089±0.00003 |
| $\frac{\Delta c}{\Delta T}$ (Å/K) | 0.00066 ± 0.00002 | 0.00100±0.00002 |

At the expanded volumes at 100, 150 and 200 K, we then calculate the ADPs again for all three levels of theory (Fig. 8). PBE+D3(BJ) now fits best to experiment; vdW-DF2 and PBE+TS show less satisfactory results than in the harmonic approximation. The latter result (which seems counterintuitive at first) emphasizes that a certain degree of error compensation may have been in play before. In turn, the underestimation of ADPs by PBE+D3(BJ) in the harmonic approximation can now partly be attributed to the neglect of thermal expansion of the crystal. Finally, all calculated ADPs still overestimate the experimental ones at 200 K. This may be explicable by the overestimated volume expansion at all levels of theory (Table 4).

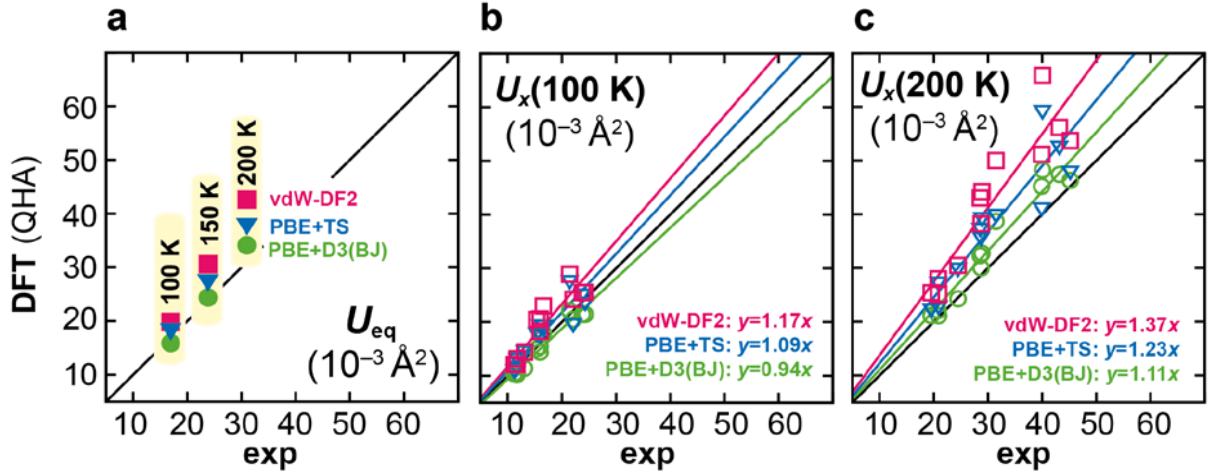

FIG. 8. (a) Equivalent displacement parameters ($U_{eq}$) of α-sulfur in the QHA. (b) and (c) Same for the main-axis components of the anisotropic displacement tensor at 100 K and 200 K. Results of a linear least-squares fit to the data points are noted.

Summarizing the results of the present study, Fig. 9 compares the calculated ADPs in the harmonic approximation and in the QHA at the best tested level of theory, the PBE+D3(BJ) level, to the experimentally derived ADPs at 100 and 200 K. Both the size and shape of ADPs in the QHA fit very well to the single-crystal XRD-data; only small differences in the main-axis components are still visible. The results of the harmonic and the quasi-harmonic approximation mainly differ in the size but not so much in the shape of the ADPs. In conclusion, the agreement of calculated ADPs with experimental ones can be improved when using the quasi-harmonic approximation instead of the purely harmonic approximation.

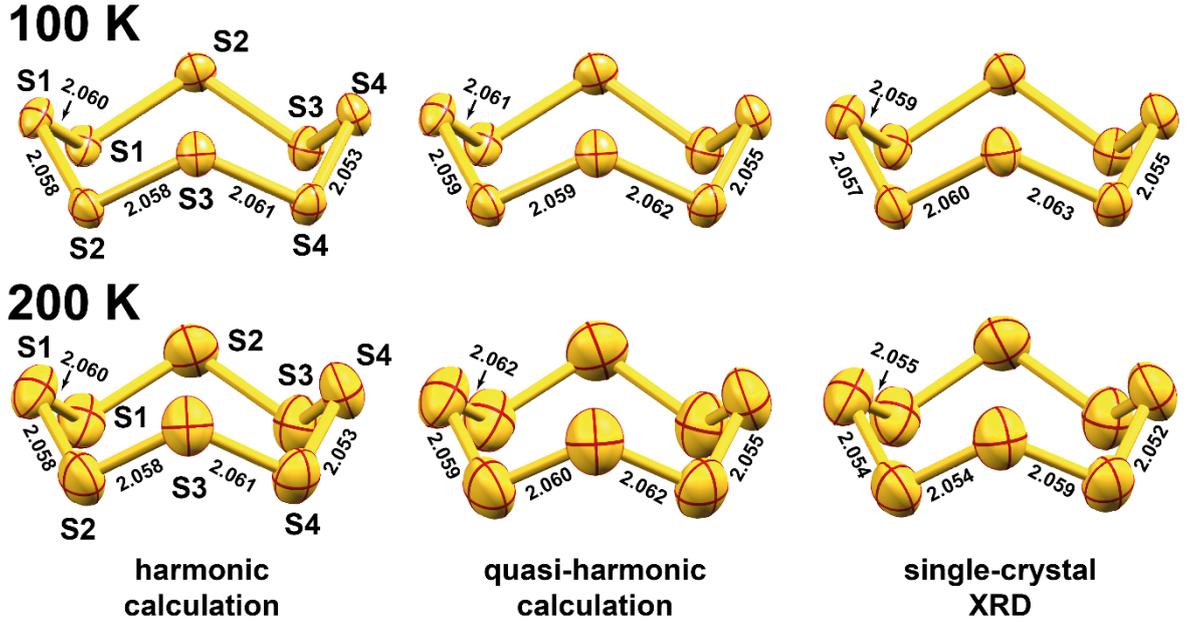

FIG. 9. Top: Displacement ellipsoids[33] for the $S_8$ ring in α-sulfur at the 90% probability level, comparing DFT-D based phonon computations (PBE+D3(BJ) level) in the harmonic (left) and in the quasi-harmonic approximation (middle) with single-crystal XRD results at standard resolution (right); all refer to a temperature of 100 K. Symmetry-inequivalent atom labels and bond lengths (in Å) are given. Bottom: Same for 200 K.

## V. CONCLUSIONS

Using a combined approach of experiment and theory, we have systematically studied the thermal expansion and motion of atoms in α-sulfur, a prototypical elemental solid. In particular, we explored how the predicted unit-cell volume and its thermal expansion affect calculated ADPs. Previous studies have seen very good agreement in the harmonic approximation between calculated and experimental ADPs, but our study emphasizes that this must sometimes be attributed to error compensation—namely, in cases where the calculated ground-state volume is strongly different from experiment. We have also seen that careful comparison between calculated and experimental vibrational frequencies can be a good indicator for the ensuing quality of finite-temperature predictions: well-suited DFT methods for ADP calculation should reproduce the experimental volumes and frequencies near zero Kelvin sufficiently well; the "best" suited combination of DFT functional and dispersion correction technique may very well depend on the specific application, and the best-performing method for α-sulfur might not be the ideal choice for materials of very different composition. That being said, we have shown that including thermal expansion in the ADP calculation can substantially improve the agreement with experiment for such methods (in the case of α-sulfur: up to 200 K). As ADPs make up for two thirds of the variables in structure determination, their replacement by theoretical values offers an efficient way to increase the ratio between observations and parameters. This improvement will be beneficial for XRD-based structure models of standard resolution, and may be downright decisive in cases where only limited diffraction information is available, e.g., in neutron and/or powder diffraction.

## SUPPLEMENTARY MATERIAL

See supplementary material for detailed information about the sample used for X-ray powder diffraction and lattice parameters and the isotropic displacement parameters resulting from the profile fitting of this data. Moreover, information on the DFT calculations is included.

## ACKNOWLEDGEMENTS

We thank Dr. Wolfram Klein, Verbio AG, for providing excellent microcrystals of α-$S_8$ for powder diffraction. Computer time was provided by the Jülich–Aachen Research Alliance (project jara0069). A Chemiefonds fellowship for J. G. and a CSC fellowship for A. W. are gratefully acknowledged.

# Supplementary Information for the Manuscript

# Lattice thermal expansion and anisotropic displacements in α-sulfur from diffraction experiments and first-principles theory

Janine George,[1] Volker L. Deringer,[1,†] Ai Wang,[1] Paul Müller,[1] Ulli Englert,[1,*] and Richard Dronskowski[1,2,*]

[1] Institute of Inorganic Chemistry, RWTH Aachen University, Landoltweg.1, Aachen 52074, Germany. E-mail: : ullrich.englert@ac.rwth-aachen.de; drons@HAL9000.ac.rwth-aachen.de

[2] Jülich-Aachen Research Alliance (JARA-HPC), RWTH Aachen University, Aachen 52056, Germany

[†] Present address: Engineering Laboratory, University of Cambridge, Trumpington Street, Cambridge CB2 1PZ, United Kingdom

# EXPERIMENTAL RESULTS

## Microcrystalline α-S$_8$ via desulphurization in a biorefinery

At its Zörbig site (near Halle in Saxonia-Anhalt, Germany), Verbio AG operates a biorefinery that converts all digestible components from crop and straw either into bioethanol or biomethane. The waste from the bioethanol process called stillage is fed into specially designed anaerobic reactors where bacteria generate crude biogas, a mixture of mainly $CH_4$, $CO_2$ and traces of $H_2S$. In the 35 MW biogas process, the first step is $H_2S$ recovery from the crude gas. This is achieved in a caustic scrubber combined with an aerobic step in the presence of special bacteria of the thiobacillus type. These species selectively convert $H_2S$ into elemental sulfur which is removed from the slurry and further treated for applications in the fertilizer industry.

## Crystal data and refinement results of single crystal X-ray diffraction

Table S1 Crystal data and refinement results of single crystal X-ray diffraction at 100 K, 150 K and 200 K. For data at 100 K, refinement results for high and for standard resolution are shown; within these resolution limits, displacement parameters do not differ significantly.

| Temperature | 100 K | 100 K | 150 K | 200 K |
|---|---|---|---|---|
| Molecular weight | 256.48 | | | |
| Crystal system | Orthorhombic | | | |
| Space group | *Fddd* | | | |
| $a$ (Å) | 10.3761(4) | 10.3761(4) | 10.4004(4) | 10.4272(4) |
| $b$ (Å) | 12.7255(5) | 12.7255(5) | 12.7658(5) | 12.8088(5) |
| $c$ (Å) | 24.427(1) | 24.427(1) | 24.459(1) | 24.491(1) |
| $V$ (Å$^3$) | 3225.4(2) | 3225.4(2) | 3247.4(2) | 3271.0(2) |
| Z | 16 | | | |
| Crystal size (mm) | 0.22×0.15×0.08 | | | |
| No. of parameters | 37 | | | |
| $2\theta$(°) | 90.4 | 61.2 | 62.0 | 61.2 |
| Total/unique reflection | 31821/3321 | 8243/1222 | 16579/1282 | 8871/1241 |
| R$_{int}$ | 0.0652 | 0.0447 | 0.0677 | 0.0498 |
| R[$F^2$>2σ($F^2$)] | 0.0324 | 0.0275 | 0.0273 | 0.0308 |
| wR$_2$($F^2$) | 0.0689 | 0.0575 | 0.0548 | 0.0601 |
| GOF | 1.053 | 1.050 | 1.121 | 1.059 |
| Δ$_{\rho max}$/Δ$_{\rho min}$ (eÅ$^{-3}$) | 0.621/-0.684 | 0.458/-0.423 | 0.378/-0.435 | 0.506/-0.336 |

We show the ADPs derived from single-crystal diffraction in Figure S1. The numerical values of $U_{eq}$ are also compared to the results from powder diffraction data in Figure S2.

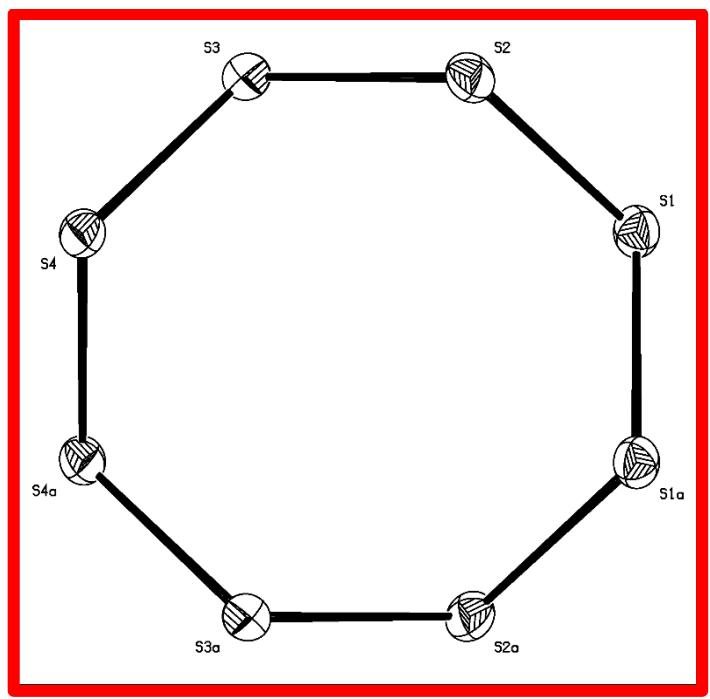

**Figure S1.** ORTEP plot (50% probability level) of α-$S_8$ at 100 K ($a=3/4-x, 3/4-y, z$).

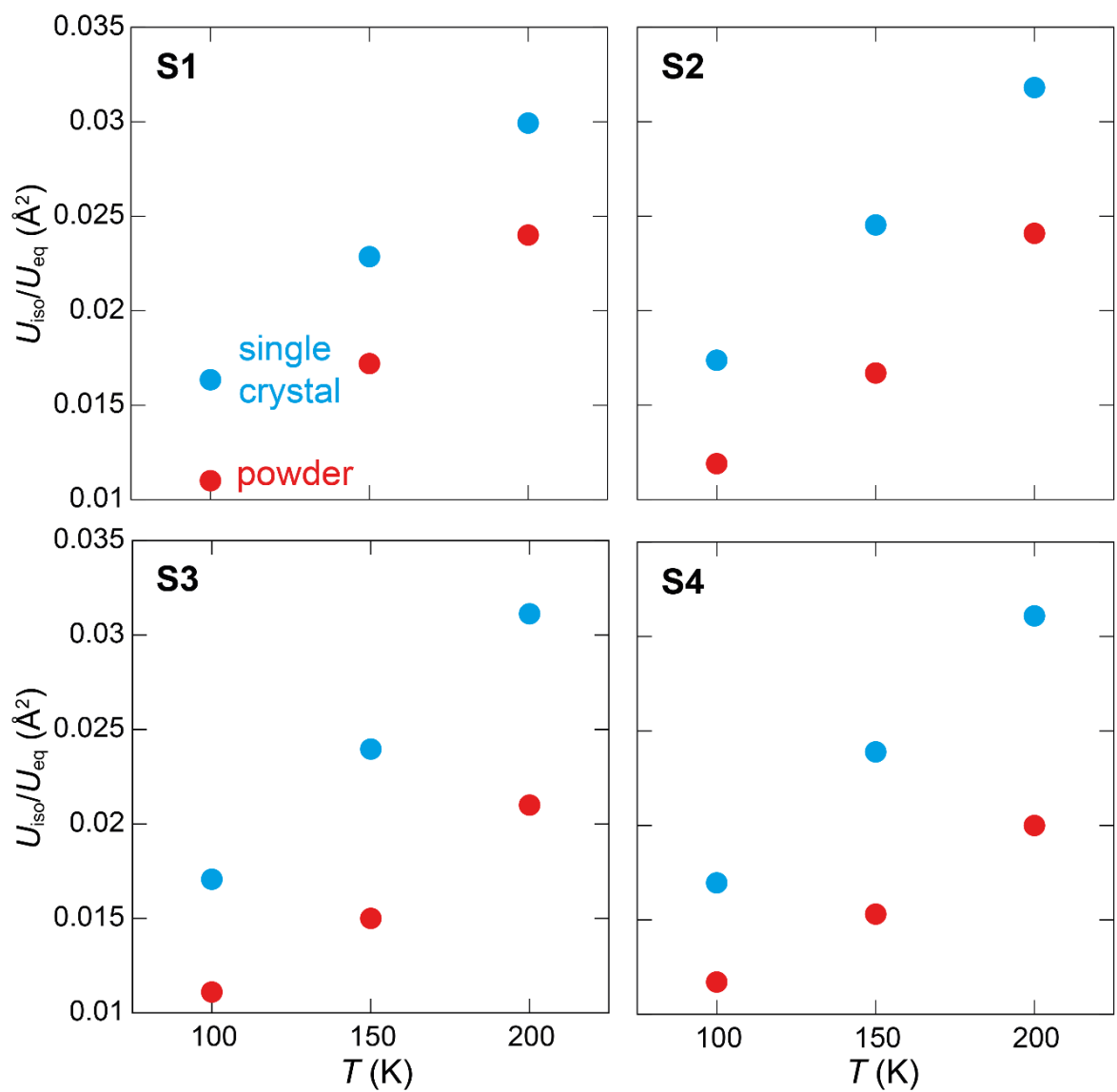

**Figure S2**. $U_{eq}/U_{iso}$ from single-crystal and powder diffraction for the crystallographically inequivalent four S atoms at 100, 150 and 200 K.

## Lattice parameters from powder diffraction data

**Table S2.** Lattice parameters from powder diffraction data.

| $T$ (K) | $a$ (Å) | $b$ (Å) | $c$ (Å) |
|---|---|---|---|
| 298 | 10.4925(4) | 12.9022(5) | 24.559(1) |
| 250 | 10.4616(4) | 12.8592(5) | 24.533(1) |
| 200 | 10.4272(4) | 12.8088(5) | 24.491(1) |
| 150 | 10.4004(4) | 12.7658(5) | 24.459(1) |
| 100 | 10.3761(4) | 12.7255(5) | 24.427(1) |
| 50 | 10.3574(4) | 12.6925(4) | 24.399(1) |
| 10 | 10.3524(4) | 12.6819(5) | 24.394(1) |

# DFT CALCULATIONS

## Ground-state

**k**-point Meshes:

**Table S3. k**-point convergence of the electronic energies for the different methods used in our calculation. The energy is obviously converged in terms of the **k**-point mesh size.

| Method | Structure | k-point mesh for the electronic structure calculation | Energy (eV) |
| --- | --- | --- | --- |
| vdW-DF2 | Fully optimized | 4×4×2 | −297.77280600 |
| vdW-DF2 | Optimized at the vdW-DF2 level with a 4×4×2 **k**-point mesh | 5×5×3 | −297.77282702 |
| PBE+D3(BJ) | Fully optimized | 4×4×2 | −553.32306545 |
| | Optimized at the PBE+D3(BJ) level with a 4×4×2 **k**-point mesh | 5×5×3 | −553.32322382 |
| PBE+TS | Fully optimized | 4×4×2 | −542.06770668 |
| PBE+TS | Optimized at the PBE+TS level with a 4×4×2 **k**-point mesh | 5×5×3 | −542.06777952 |

Calculated ADPs

**q-Point Meshes (for the ADP Calculation within Phonopy)**

We show the convergence of the *q*-point mesh for the ADPs calculated at 300 K because ADPs at higher temperatures are more difficult to converge. We used a frequency-cutoff of 0.1 THz for the ADP calculation with Phonopy. The given $U_{eq}$ is an arithmetic mean of all 4 sulfur atoms as defined in the main text.

**Table S4.** Convergence of the ADPs when a higher q-point mesh is applied. The ADPs are converged at a 52×52×26 q-point mesh. We use this mesh for all other ADP calculations.

| Method | q-point mesh | $U_{eq}$ (Å$^3$) |
|---|---|---|
| PBE+D3(BJ) | 40×40×20 | 0.03872 |
| PBE+D3(BJ) | 44×44×22 | 0.03870 |
| PBE+D3(BJ) | 48×48×24 | 0.03869 |
| PBE+D3(BJ) | 52×52×26 | 0.03870 |
| PBE+TS | 48×48×24 | 0.04486 |
| PBE+TS | 52×52×26 | 0.04486 |
| PBE+TS | 60×60×30 | 0.04487 |
| vdW-DF2 | 50×50×25 | 0.04639 |
| vdW-DF2 | 52×52×26 | 0.04641 |

Equivalent displacement parameters ($U_{eq}$) of α-sulfur in the harmonic approximation

**Table S5.** Equivalent displacement parameters $U_{eq}$ (Å$^2$) calculated in the harmonic approximation

| T(K) | vdW-DF2 | PBE+TS | PBE+D3(BJ) |
|---|---|---|---|
| 100 | 0.01643 | 0.01588 | 0.01383 |
| 150 | 0.02378 | 0.02299 | 0.01991 |
| 200 | 0.03127 | 0.03023 | 0.02612 |

Main-axis components $U_1$, $U_2$, $U_3$ of α-sulfur in the harmonic approximation

**Table S6.** Main-axis components $U_x$ (Å$^2$) calculated in the harmonic approximation.

| Atom | vdW-DF2 | | | PBE+TS | | | PBE+D3(BJ) | | |
|---|---|---|---|---|---|---|---|---|---|
| **100 K** | | | | | | | | | |
| S1 | 0.01027 | 0.01539 | 0.02233 | 0.01020 | 0.01535 | 0.02208 | 0.00887 | 0.01236 | 0.01888 |
| S2 | 0.01085 | 0.01861 | 0.02139 | 0.01177 | 0.01688 | 0.02045 | 0.00920 | 0.01517 | 0.01893 |
| S3 | 0.01182 | 0.01723 | 0.02021 | 0.01224 | 0.01655 | 0.01815 | 0.01005 | 0.01319 | 0.01813 |
| S4 | 0.01033 | 0.01723 | 0.02144 | 0.01014 | 0.01699 | 0.01982 | 0.00905 | 0.01292 | 0.01917 |
| **200 K** | | | | | | | | | |
| S1 | 0.01904 | 0.02914 | 0.04304 | 0.01893 | 0.02912 | 0.04258 | 0.01629 | 0.02316 | 0.03616 |
| S2 | 0.02017 | 0.03566 | 0.04112 | 0.02203 | 0.03221 | 0.03933 | 0.01690 | 0.02891 | 0.03618 |
| S3 | 0.02208 | 0.03292 | 0.03877 | 0.02293 | 0.03162 | 0.03472 | 0.01855 | 0.02494 | 0.03465 |
| S4 | 0.01913 | 0.03293 | 0.04121 | 0.01880 | 0.03248 | 0.03802 | 0.01661 | 0.02440 | 0.03669 |

# Optimized crystal structures in POSCAR format

```
PBE+D3BJ
  1.00000000000000
    10.2925136190973632     0.0000000747456260     0.0000002086431546
    -0.0000000843121777    12.6479877392430176    -0.0000001290916734
     0.0000005371750222    -0.0000002503479979    24.3177278060198248
   S
  128
Direct
  0.8565721507888213  0.9547673957859573  0.9492590711989664
  0.1434278592281117  0.0452326044622708  0.0507409300553263
  0.8934278623957184  0.7952325815328010  0.9492590743870508
  0.1065721474988877  0.2047674186805182  0.0507409259978573
  0.8934278648424367  0.9547673948381430  0.8007409311267608
  0.1065721372460402  0.0452326040534459  0.1992590696832508
  0.8565721531438015  0.7952325806468821  0.8007409247394932
  0.1434278490956586  0.2047674182223815  0.1992590760652178
  0.8565721418624435  0.4547673967101673  0.4492590707066650
  0.1434278482751736  0.5452326038530799  0.5507409295881374
  0.8934278536379523  0.2952325824717192  0.4492590747608602
  0.1065721364244183  0.7047674180580756  0.5507409264044725
  0.8934278674649505  0.4547673963383545  0.3007409313400515
  0.1065721304475389  0.5452326052227932  0.6992590694479901
  0.8565721557074824  0.2952325821458572  0.3007409245149830
  0.1434278420780615  0.7047674194130948  0.6992590762183113
  0.3565721418367644  0.9547673965860994  0.4492590708996715
  0.6434278478670947  0.0452326036887243  0.5507409295519849
  0.3934278537619207  0.7952325823936235  0.4492590748724368
  0.6065721360844023  0.2047674178993617  0.5507409264790866
  0.3934278670486364  0.9547673965338390  0.3007409313812701
  0.6065721305399734  0.0452326052876586  0.6992590694307523
  0.3565721556007944  0.7952325823035267  0.3007409246209107
  0.6434278422889150  0.2047674194503983  0.6992590761772277
  0.3565721507848778  0.4547673958249234  0.9492590717244642
  0.6434278590307585  0.5452326044421767  0.0507409299385415
  0.3934278626452041  0.2952325816146555  0.9492590744106622
  0.6065721472453873  0.7047674186545763  0.0507409259610085
  0.3934278647330487  0.4547673948521194  0.8007409312116707
  0.6065721376474116  0.5452326038881452  0.1992590696696510
  0.3565721532186359  0.2952325806370979  0.8007409248510200
  0.6434278491846612  0.7047674180981360  0.1992590759610593
  0.7837690843981662  0.0351698406786198  0.0762142664133663
  0.2162309249200121  0.9648301566199109  0.9237857352050440
  0.9662309158777092  0.7148301581244638  0.0762142668048256
  0.0337690931884467  0.2851698391013144  0.9237857332022728
  0.9662309101694859  0.0351698528369226  0.6737857425647107
  0.0337690860783439  0.9648301477840633  0.3262142580333887
  0.7837690785413969  0.7148301704250031  0.6737857256592221
  0.2162309178270476  0.2851698301129488  0.3262142654489853
  0.7837690737349305  0.5351698432908378  0.5762142656348317
  0.2162309174883603  0.4648301598940137  0.4237857344726521
  0.9662309053131182  0.2148301608281287  0.5762142675565443
  0.0337690856841633  0.7851698423178846  0.4237857338866959
  0.9662309182714566  0.5351698539429961  0.1737857431431983
  0.0337690852500714  0.4648301457481097  0.8262142577958684
  0.7837690865183617  0.2148301716483800  0.1737857250571722
  0.2162309167701579  0.7851698281810044  0.8262142755129105
  0.2837690732944225  0.0351698432591974  0.5762142657126432
  0.7162309170266568  0.9648301602063540  0.4237857343128084
  0.4662309054448599  0.7148301609017196  0.5762142676686963
  0.5337690852065720  0.2851698426163196  0.4237857340622995
  0.4662309176411625  0.0351698539694638  0.1737857431216980
  0.5337690857989443  0.9648301460327673  0.8262142578526905
  0.2837690864455240  0.7148301714423013  0.1737857252357244
  0.7162309172520764  0.2851698284238395  0.8262142754070041
  0.2837690844151055  0.5351698404212826  0.0762142665312737
  0.7162309246342105  0.4648301569248616  0.9237857350175744
  0.4662309164086977  0.2148301580697805  0.0762142667864012
  0.5337690926402772  0.7851698393446966  0.9237857332636850
  0.4662309101572859  0.5351698529873588  0.6737857426902707
  0.5337690867006160  0.4648301480084527  0.3262142581510545
  0.2837690787774605  0.2148301704778675  0.6737857257371260
  0.7162309180370912  0.7851698305263781  0.3262142751221049
  0.7057590305571395  0.9856807875180849  0.0026593103544599
  0.2942409791073146  0.0143192119593749  0.9973406907863591
  0.0442409797731926  0.7643191993492948  0.0026593097311363
  0.9557590297037422  0.2356807999178443  0.9973406907393780
  0.0442409800262809  0.9856807897974917  0.7473406905342372
  0.9557590197130921  0.0143192093615170  0.2526593091536000
  0.7057590306735264  0.7643192020335476  0.7473406908220284
  0.2942409691920744  0.2356807970100263  0.2526593111161617
  0.7057590210590092  0.4856807889562589  0.5026593095898377
  0.2942409696442780  0.5143192124010838  0.4973406909577349
  0.0442409703745312  0.2643192009858097  0.5026593104912038
  0.9557590201445763  0.7356808002012301  0.4973406904937363
  0.0442409847966516  0.4856807914105588  0.2473406918078140
  0.9557590153514539  0.5143192098671392  0.7526593100410182
  0.7057590353740508  0.2643192037753366  0.2473406895355694
  0.2942409645708466  0.7356807977025852  0.7526593101301202
  0.2057590208310955  0.9856807887224193  0.5026593096568845
  0.7942409691875341  0.0143192123450788  0.4973406907887652
  0.5442409705248252  0.7643192010672308  0.5026593106231303
  0.4557590197519090  0.2356808002348743  0.4973406906239788
  0.5442409842862901  0.9856807917656951  0.2473406918307859
  0.4557590156210622  0.0143192099837819  0.7526593101224677
  0.2057590352679526  0.7643192036896735  0.2473406896802501
  0.7942409649340263  0.2356807976880120  0.7526593100426595
  0.2057590305459200  0.4856807873408400  0.0026593104566146
  0.7942409788502118  0.5143192120995792  0.9973406906298834
  0.5442409801303825  0.2643191995888401  0.0026593097419649
  0.4557590293291440  0.7356807999109805  0.9973406907846041
  0.5442409799828312  0.4856807899364739  0.7473406906560882
  0.4557590202330815  0.5143192091655280  0.2526593092108129
  0.2057590308514250  0.2643192019274565  0.7473406909087004
  0.7942409693725025  0.7356807971665233  0.2526593109284079
  0.7849163615794978  0.9104622601389565  0.1305330575490586
  0.2150836462057697  0.0895377357258624  0.8694669431275059
  0.9650836463961809  0.8395377504714219  0.1305330524035142
  0.0349163613885253  0.1604622453618774  0.8694669485116506
  0.9650836354892078  0.9104622767938295  0.6194669458774200
  0.0349163568310757  0.0895377248577489  0.3805330540903213
  0.7849163507155907  0.8395377672691566  0.6194669455957751
  0.2150836416380670  0.1604622343385245  0.3805330560395461
  0.7849163505090857  0.4104622635049253  0.6305330572833299
  0.2150836421319369  0.5895377409067777  0.3694669436347269
  0.9650836353501546  0.3395377538660540  0.6305330526266264
  0.0349163572955717  0.6604622505318218  0.3694669479423851
  0.9650836468177530  0.4104622775273654  0.1194669466415732
  0.0349163606217004  0.5895377210646586  0.8805330549620862
  0.7849163620062782  0.3395377680714660  0.1194669447826726
  0.2150836453842686  0.6604622306150674  0.8805330570390396
  0.2849163503460872  0.9104622635197614  0.6305330574097994
  0.7150836417096897  0.0895377414698331  0.3694669436220153
  0.4650836352363825  0.8395377539943283  0.6305330527677597
  0.5349163568548860  0.1604622510577443  0.3694669479776778
  0.4650836465390142  0.9104622773595352  0.1194669467613991
  0.5349163611354513  0.0895377214246054  0.8805330549638271
  0.2849163618351227  0.8395377677340363  0.1194669448946541
  0.7150836458544916  0.1604622310318717  0.8805330550779047
  0.2849163617641821  0.4104622598379208  0.1305330576400294
  0.7150836458351719  0.5895377362012368  0.8694669430462909
  0.4650836466429382  0.3395377503057517  0.1305330524639103
  0.5349163609506675  0.6604622457513685  0.8694669484947468
  0.4650836355179990  0.4104622769782509  0.6194669459826230
  0.5349163572874289  0.5895377253580207  0.3805330540410452
  0.2849163507958750  0.3395377673604401  0.6194669457204398
  0.7150836420034707  0.6604622349517086  0.3805330559565974
```

```
vdW-DF2
   1.00000000000000
     10.7830322260653837    0.0000001183341628    0.0000004658542724
     -0.0000001522965020   13.1409896683133969    0.0000002073475172
      0.0000012040396102    0.0000008356044973   25.3304854792396092
   S
   128
Direct
  0.8575319816699718  0.9536238636187520  0.9507030132808865
  0.1424680324839684  0.0463761349598570  0.0492969877432756
  0.8924680090458708  0.7963761337319539  0.9507030271214205
  0.1075320052349795  0.2036238652327498  0.0492969729773876
  0.8924680391907387  0.9536238887369350  0.7992969699828905
  0.1075319653329956  0.0463761097582989  0.2007030307488264
  0.8575320109900701  0.7963761587364147  0.7992969895684396
  0.1424679933993005  0.2036238398462586  0.2007030114886845
  0.8575319694615118  0.4536238673668294  0.4507030132694183
  0.1424680155125131  0.5463761340411821  0.5492969868566817
  0.8924679961291062  0.2963761372430227  0.4507030274988182
  0.1075319877421919  0.7036238644150430  0.5492969736248270
  0.8924680426180984  0.4536238916860214  0.2992969870386713
  0.1075319535144317  0.5463761098949149  0.7007030301725621
  0.8575320147957939  0.2963761616171041  0.2992969887994832
  0.1424679806874565  0.7036238395900654  0.7007030116773265
  0.3575319696257679  0.9536238668735280  0.4507030138824106
  0.6424680146765667  0.0463761340015765  0.5492969863033821
  0.3924679974963041  0.7963761370595321  0.4507030281665578
  0.6075319868381968  0.2036238645710284  0.5492969734955366
  0.3924680409805816  0.9536238925111817  0.2992969710128150
  0.6075319547093372  0.0463761100178317  0.7007030300081993
  0.3575320136740956  0.7963761622727716  0.2992969899256713
  0.6424679820709400  0.2036238398025176  0.7007030109255794
  0.3575319829441881  0.4536238645252908  0.9507030146411353
  0.6424680328598953  0.5463761357514798  0.0492969870331734
  0.3924680110297132  0.2963761345127480  0.9507030278783546
  0.6075320045661243  0.7036238655188427  0.0492969725061201
  0.3924680385035515  0.4536238880764003  0.7992969706344155
  0.6075319662781169  0.5463761094812867  0.2007030305113346
  0.3575320109958540  0.2963761579209176  0.7992969899670967
  0.6424679942182081  0.7036238403462960  0.2007030107501890
  0.7847974834830822  0.0331904364646221  0.0755713123084063
  0.2152025296130660  0.9668095589132122  0.9244286902627223
  0.9652025091795124  0.7168096001188005  0.0755713234394264
  0.0347975054502001  0.2831903958838566  0.9244286762819200
  0.9652024890393562  0.0331903998000627  0.6744286546666132
  0.0347975028150742  0.9668096005632094  0.3255713451938647
  0.7847974633174957  0.7168095631308447  0.6744287106891989
  0.2152025290421946  0.2831904369394991  0.3255712912289894
  0.7847974682673424  0.5331904395823983  0.5755713110310268
  0.2152025191226841  0.4668095647392860  0.4244286889633457
  0.9652024918042628  0.2168096022158750  0.5755713239212739
  0.0347974943412410  0.7831904017345153  0.4244286772539522
  0.9652025040273529  0.5331904038372528  0.1744286561026129
  0.0347975029301750  0.4668095956300746  0.8255713449621922
  0.7847974793156069  0.2168095669522927  0.1744287087760199
  0.2152025272098612  0.7831904325128889  0.8255712913099558
  0.2847974678653813  0.0331904386877824  0.5755713116496324
  0.7152025154192501  0.9668095669640380  0.4244286877951851
  0.4652024932402199  0.7168096023473112  0.5755713246785774
  0.5347974914400453  0.2831904040977804  0.4244286779952091
  0.4652025013174850  0.0331904019989508  0.1744286563981987
  0.5347975049650415  0.9668095980331159  0.8255713453742146
  0.2847974764324519  0.7168095649460327  0.1744287103678275
  0.7152025307363132  0.2831904345886684  0.8255712899082397
  0.2847974868564407  0.5331904346211616  0.0755713131832962
  0.7152025283243617  0.4668095610644372  0.9244286890137587
  0.4652025122861119  0.2168095981911051  0.0755713235370052
  0.5347975014629611  0.7831903970776892  0.9244286762719298
  0.4652024886989068  0.5331903995958314  0.6744286554352428
  0.5347975070834963  0.4668096021219981  0.3255713453859030
  0.2847974634042671  0.2168095627710471  0.6744287109548068
  0.7152025315733042  0.7831904397123637  0.3255712896825855
  0.7096181939943591  0.9852674462785842  0.0029575239883428
  0.2903818195810643  0.0147325512912317  0.9970424767654933
  0.0403818012261112  0.7647325776176785  0.0029575286834671
  0.9596182133746254  0.2352674205833125  0.9970424717352486
  0.0403818081727181  0.9852674431975075  0.7470424436300362
  0.9596181913434094  0.0147325558120599  0.2529575544842260
  0.7096182004912848  0.7647325748120508  0.7470425035612465
  0.2903817989330904  0.2352674240746637  0.2529574999644026
  0.7096181804654478  0.4852674504565044  0.5029575238366633
  0.2903818050935101  0.5147325526424353  0.4970424775785673
  0.0403817864112384  0.2647325803999223  0.5029575291617618
  0.9596181983800349  0.7352674216591026  0.4970424707944829
  0.0403818160356408  0.4852674468386695  0.2470424467341701
  0.9596181839221742  0.5147325538245298  0.7529575566449012
  0.7096182090350140  0.2647325778753711  0.2470425000211875
  0.2903817908332869  0.7352674230762943  0.7529574977991942
  0.2096181803360011  0.9852674492290561  0.5029575246482665
  0.7903818030934815  0.0147325529985594  0.4970424761066354
  0.5403817878044066  0.7647325811361014  0.5029575298275972
  0.4596181969019426  0.2352674220032185  0.4970424715761794
  0.5403818141649452  0.9852674473842029  0.2470424463883987
  0.4596181857321326  0.0147325546887984  0.7529575564616096
  0.2096182075674591  0.7647325778751224  0.2470425023729703
  0.7903817931660768  0.2352674237173815  0.7529574968164852
  0.2096181960781962  0.4852674462570690  0.0029575260757824
  0.7903818194259316  0.5147325527774669  0.9970424755417326
  0.5403818035298116  0.2647325781570871  0.0029575287672614
  0.4596182110593361  0.7352674203968874  0.9970424717085962
  0.5403818076121070  0.4852674431291888  0.7470424442454089
  0.4596181939237098  0.5147325555425013  0.2529575550948877
  0.2096182006870038  0.2647325739484856  0.7470425041357487
  0.7903818001179701  0.7352674251425384  0.2529574980984677
  0.7865960563922769  0.9094136288495989  0.1285107820584557
  0.2134039554000466  0.0905863654819612  0.8714892182575795
  0.9634039098837570  0.8405864282968167  0.1285107733228301
  0.0365960715324078  0.1594135662657337  0.8714892273788664
  0.9634039082608510  0.9094135630682914  0.6214892102865406
  0.0365960775592171  0.0905864382085610  0.3785107889844994
  0.7865960240811560  0.8405863623353582  0.6214892341093332
  0.2134039618840973  0.1594136385134064  0.3785107683705675
  0.7865960390862057  0.4094136310029199  0.6285107818856517
  0.2134039501877893  0.5905863742632036  0.3714892192720072
  0.9634039230505493  0.3405864296465069  0.6285107737262052
  0.0365960664717022  0.6594135746385419  0.3714892263027920
  0.9634039294542163  0.4094135676987136  0.1214892117524542
  0.0365960836453425  0.5905864308087558  0.8785107903847873
  0.7865960453720291  0.3405863666118734  0.1214892322304379
  0.2134039680293469  0.6594136324732389  0.8785107666535197
  0.2865960391865485  0.9094136301075579  0.6285107826282470
  0.7134039464393922  0.0905863778281457  0.3714892185026386
  0.4634039236036642  0.8405864292834977  0.6285107745400680
  0.5365960630190116  0.1594135785561335  0.3714892260951572
  0.4634039260495655  0.9094135637624063  0.1214892123839704
  0.5365960858236818  0.0905864343989222  0.8785107897986677
  0.2865960420270213  0.8405863629552712  0.1214892335683260
  0.7134039705253699  0.1594136351590620  0.8785107660247320
  0.2865960598868114  0.4094136254123768  0.1285107833909507
  0.7134039530739571  0.5905863680402632  0.8714892175892928
  0.4634039441045843  0.3405864248433588  0.1285107742460383
  0.5365960683081852  0.6594135682322175  0.8714892270305512
  0.4634039080553478  0.4094135627610527  0.6214892109075620
  0.5365960818109485  0.5905864418233193  0.3785107887937329
  0.2865960238748428  0.3405863621448972  0.6214892346912393
  0.7134039649706310  0.6594136421477543  0.3785107674731805
```

**PBE+TS**
   1.00000000000000
     10.7178303534345485    0.0000000787330541    0.0000001754394700
    -0.0000000937559934   13.0810490583603158   -0.0000006663194022
     0.0000000819427089   -0.0000010220174998   25.2148615044681748

   S
   128
Direct
  0.8570086243986736  0.9521692032324864  0.9534633351244395
  0.1429914089420734  0.0478307999771346  0.0465366654345800
  0.8929913412066384  0.7978309073723437  0.9534632912870507
  0.1070086950203901  0.2021690967335346  0.0465367070124287
  0.8929913557836002  0.9521691446331744  0.7965366962775917
  0.1070086415305411  0.0478308520007147  0.2034633033592570
  0.8570086386843769  0.7978308487439918  0.7965366711084485
  0.1429913588819289  0.2021691479012162  0.2034633307974758
  0.8570085893053729  0.4521692059541920  0.4534633355794142
  0.1429913703825463  0.5478307948306735  0.5465366580766613
  0.8929913016770286  0.2978309078905568  0.4534632938565082
  0.1070086615954082  0.7021690937703795  0.5465367030319115
  0.8929913823287023  0.4521691544569038  0.2965366966571281
  0.1070086160397068  0.5478308501841838  0.7034633051639858
  0.8570086676365563  0.2978308580100872  0.2965366697480576
  0.1429913274123464  0.7021691477786831  0.7034633336196592
  0.3570085990457557  0.9521691995103936  0.4534633378869728
  0.6429913659722644  0.0478307980706774  0.5465366636509970
  0.3929913202707169  0.7978309036098423  0.4534632948761583
  0.6070086455027663  0.2021690923522215  0.5465367100707894
  0.3929913692437523  0.9521691521139459  0.2965367007303001
  0.6070086196706086  0.0478308511652656  0.7034633009596334
  0.3570086551397793  0.7978308558598783  0.2965366737474326
  0.6429913345140648  0.2021691457663906  0.7034633250965783
  0.3570086310423690  0.4521692001616913  0.9534633334036266
  0.6429914113528881  0.5478308014265281  0.0465366635340700
  0.3929913446362008  0.2978309030678332  0.9534632986939902
  0.6070086965718318  0.7021690990943483  0.0465367030088473
  0.3929913570274692  0.4521691391554725  0.7965366971230807
  0.6070086527361838  0.5478308524874436  0.2034633050360029
  0.3570086468703408  0.2978308418961433  0.7965366673646983
  0.6429913609316387  0.7021691508908035  0.2034633308945644
  0.7854887730957643  0.0305468202682156  0.0763331922044799
  0.2145112582858815  0.9694531913670446  0.9236668100311789
  0.9645113536283247  0.7194530919054500  0.0763330820291657
  0.0354886852053653  0.2805469035461030  0.9236669146510224
  0.9645112906549826  0.0305468796671704  0.6736668774597661
  0.0354886806609116  0.9694531135473738  0.3263331188473941
  0.7854887127457033  0.7194531699612412  0.6736668422189283
  0.2145112641547300  0.2805468204227779  0.3263331637453604
  0.7854887453188937  0.5305468109443154  0.5763331926435526
  0.2145112282950237  0.4694531991742821  0.4236668007930717
  0.9645113110051398  0.2194530959977996  0.5763330845800070
  0.0354886634040881  0.7805469140939252  0.4236669121643715
  0.9645113267428229  0.5305468933622564  0.1736668792362792
  0.0354886918703912  0.4694531028411717  0.8263331229495279
  0.7854887543273179  0.2194531811271574  0.1736668386267368
  0.2145112562321643  0.7805468172927874  0.8263331642095650
  0.2854887383547364  0.0305467966293804  0.5763331951059598
  0.7145112205504631  0.9694532118368855  0.4236668045400194
  0.4645113218914432  0.7194530883973087  0.5763330855057589
  0.5354886354476562  0.2805469208050653  0.4236669223966558
  0.4645113161641561  0.0305468867515017  0.1736668829526238
  0.5354886942651689  0.9694531197321368  0.8263331217308334
  0.2854887434342004  0.7194531747786215  0.1736668439147024
  0.7145112776561859  0.2805468291591851  0.8263331521673578
  0.2854887900766059  0.5305467968011328  0.0763332019310710
  0.7145112487165832  0.4694531900945975  0.9236668065346407
  0.4645113594720556  0.2194530848015006  0.0763330884543194
  0.5354886727788255  0.7805469008456001  0.9236669121638457
  0.4645112860839831  0.5305468902730652  0.6736668804742791
  0.5354887107664794  0.4694531184129573  0.3263331235507749
  0.2854887199734222  0.2194531756183693  0.6736668364462943
  0.7145112713935049  0.7805468339855040  0.3263331603742756
  0.7120388169607281  0.9834382258403664  0.0048891615353952
  0.2879612158628930  0.0165617739587844  0.9951108388810894
  0.0379612190748233  0.7665618117330837  0.0048890641679691
  0.9620388161323987  0.2334381905369725  0.9951109328149386
  0.0379611977163350  0.9834382235783181  0.7451108986890915
  0.9620387882313821  0.0165617727744731  0.2548890960909134
  0.7120387963536530  0.7665618101354994  0.7451108682142120
  0.2879611949314835  0.2334381825656706  0.2548891392036481
  0.7120387848787502  0.4834382319982709  0.5048891627415912
  0.2879611820740635  0.5165617718338851  0.4951108328369500
  0.0379611787763068  0.2665618105700815  0.5048890661181886
  0.9620387887361659  0.7334381941081034  0.4951109284017861
  0.0379612283231765  0.4834382358693432  0.2451109012904169
  0.9620387787107276  0.5165617644213327  0.7548891038586660
  0.7120388284520303  0.2665618200747488  0.2451108649801199
  0.2879611718927819  0.7334381855295291  0.7548891355341283
  0.2120387872785372  0.9834382188688622  0.5048891630783316
  0.7879611735426977  0.0165617809534169  0.4951108377265996
  0.5379611935394308  0.7665618103239353  0.5048890696113588
  0.4620387685031844  0.2334381894938176  0.4951109373284552
  0.5379612159816034  0.9834382344563011  0.2451109055592653
  0.4620387813985545  0.0165617728691174  0.7548891011711305
  0.2120388181130792  0.7665618175324340  0.2451108683220866
  0.7879611873943944  0.2334381844849176  0.7548891248940777
  0.2120388278654914  0.4834382239695998  0.0048891719456847
  0.7879612146404469  0.5165617748117555  0.9951108346473205
  0.5379612243707896  0.2665618066142130  0.0048890706091100
  0.4620388126055062  0.7334381903344749  0.9951109317709026
  0.5379611962614774  0.4834382272245250  0.7451108993428690
  0.4620388081781712  0.5165617678813916  0.2548891011196943
  0.2120388013930068  0.2665618054079459  0.7451108645778106
  0.7879611976081975  0.7334381933851972  0.2548891358566863
  0.7882530404927479  0.9090516175739509  0.1283269387659303
  0.2117469886156229  0.0909483705904606  0.8716730599738014
  0.9617470898695757  0.8409481699798746  0.1283269117641908
  0.0382529419475901  0.1590518217947121  0.8716730872392020
  0.9617470597516  0.9090517581891504  0.6216730779496586
  0.0382529256894628  0.0909482317755703  0.3783269203573383
  0.7882529901223947  0.8409483082808222  0.6216730651591647
  0.2117469741495270  0.1590516797291102  0.3783269389684278
  0.7882530103360068  0.4090516290481006  0.6283269405671064
  0.2117469702295693  0.5909483846419263  0.3716730555843881
  0.9617470563860451  0.3409481757506896  0.6283269124712376
  0.0382529259935609  0.6590518395359553  0.3716730818131282
  0.9617470816278555  0.4090517728496081  0.1216730787336644
  0.0382529463651338  0.5909482175143381  0.8783269258624387
  0.7882530340746925  0.3409483221221450  0.1216730626818077
  0.2117469917648258  0.6590516723337814  0.8783269374272820
  0.2882530049976069  0.9090516122551406  0.6283269413795765
  0.7117469569165848  0.0909484029503531  0.3716730614164305
  0.4617470542098729  0.8409481647124437  0.6283269159998497
  0.5382529063940424  0.1590518495163025  0.3716730878617440
  0.4617470744008543  0.9090517647882308  0.1216730835440316
  0.5382529547222603  0.0909482428657000  0.8783269200932935
  0.2882530273941626  0.8409483132096227  0.1216730662216037
  0.7117470037416993  0.1590516907747457  0.8783269303648993
  0.2882530552771314  0.4090516128904795  0.1283269477319067
  0.7117469753021624  0.5909483683248595  0.8716730569342133
  0.4617471009538363  0.3409481594693844  0.1283269191862715
  0.5382529287271183  0.6590518183360051  0.8716730858527058
  0.4617470344324488  0.4090517759881109  0.6216730770067116
  0.5382529483385099  0.5909482410893219  0.3783269216725103
  0.2882529879932889  0.3409483226094068  0.6216730636960648
  0.7117469917719674  0.6590516975283478  0.3783269379528988

## Experimental Lattice:

We used the following lattice parameters as experimental lattice parameters:

$$a=10.3199 \text{ Å}, b=12.6618 \text{ Å}, c=24.316 \text{ Å}$$

These lattice parameters are based on the original single-crystal refinement where the lattice parameters were not adapted to the more reliable powder diffraction data. Thereby, reflections were integrated with SAINT+[1] and multi-scan absorption corrections were applied with SADABS[2]. The atoms were assigned anisotropic displacement parameters, and refinements were accomplished with full-matrix least-square procedures based on $F^2$ as implemented in SHELXL-14.[3]

Main-axis components $U_1$, $U_2$, $U_3$:

**Table S7.** Main-axis components $U_x$ (Å$^2$) calculated at the experimental lattice parameters.

| Atom | vdW-DF2 | | | PBE+TS | | | PBE+D3(BJ) | | |
|---|---|---|---|---|---|---|---|---|---|
| **100 K** | | | | | | | | | |
| S1 | 0.00647 | 0.00890 | 0.01385 | 0.00755 | 0.01070 | 0.01535 | 0.00893 | 0.01243 | 0.01902 |
| S2 | 0.00674 | 0.01057 | 0.01461 | 0.00778 | 0.01266 | 0.01429 | 0.00922 | 0.01514 | 0.01917 |
| S3 | 0.00761 | 0.00935 | 0.01333 | 0.00830 | 0.01159 | 0.01302 | 0.01017 | 0.01318 | 0.01833 |
| S4 | 0.00668 | 0.00949 | 0.01449 | 0.00768 | 0.01115 | 0.01404 | 0.00900 | 0.01296 | 0.01965 |

## Quasi-harmonic approximation:

Volume expansion

**Table S8.** Course of the volume within the quasi-harmonic approximation for each DFT+D method.

| $T$ (K) | $V$ (Å$^3$/atom) | | |
|---|---|---|---|
| | vdW-DF2 | PBE+TS | PBE+D3(BJ) |
| 0 | 28.57 | 28.12 | 25.23 |
| 25 | 28.58 | 28.13 | 25.24 |
| 50 | 28.65 | 28.20 | 25.30 |
| 75 | 28.73 | 28.28 | 25.38 |
| **100** | **28.83** | **28.37** | **25.48** |
| 125 | 28.94 | 28.47 | 25.59 |
| **150** | **29.05** | **28.58** | **25.70** |
| 175 | 29.18 | 28.69 | 25.82 |
| **200** | **29.31** | **28.80** | **25.95** |
| 225 | 29.44 | 28.93 | 26.08 |
| 250 | 29.59 | 29.05 | 26.23 |
| 275 | 29.74 | 29.18 | 26.38 |
| 300 | 29.90 | 29.32 | 26.54 |
| 325 | 30.07 | 29.47 | 26.70 |
| 350 | 30.24 | 29.62 | 26.88 |

Comparison of experimental and theoretical volumetric thermal expansion coefficients

We fitted the theoretical volume-temperature curves by a polynomial of the following form: $V(T) = a + b \cdot T^2 + c \cdot T^3 + d \cdot T^4 + e \cdot T^5 + f \cdot T^6 + g \cdot T^7$ and evaluated the following equation $\alpha_V = \frac{1}{V}\left(\frac{\delta V}{\delta T}\right)_p$. For the experimental volume-temperature curves, we used the following polynomial: $V(T) = a + b \cdot T^2 + c \cdot T^3 + d \cdot T^4$.

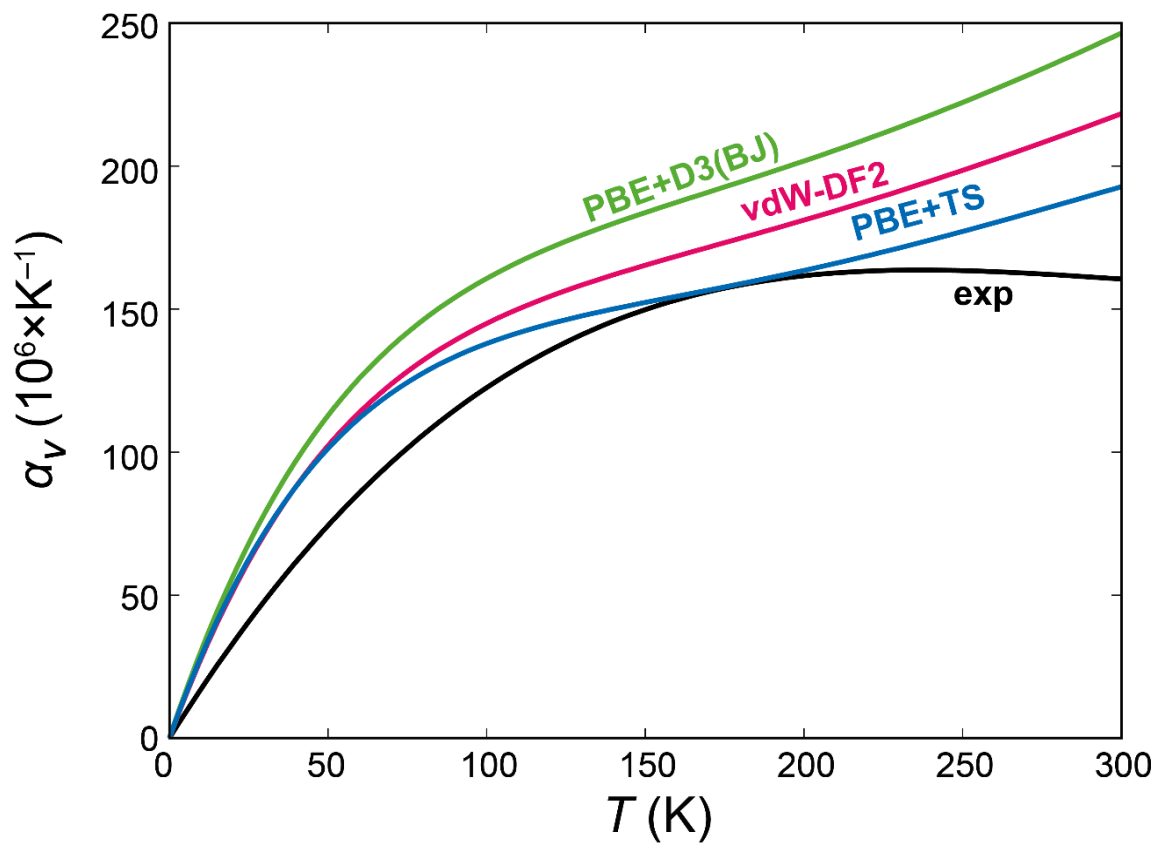

**Figure S3.** Volumetric thermal expansion coefficients for the thermal expansion derived by QHA at different levels of theory compared to the experimental results. The coefficients of the ab initio calculations are always larger or equal to the experimental ones and the deviations grow more strongly at higher temperatures.

Lattice parameters at the PBE+D3(BJ) level of theory
**Table S9.** Lattice parameters at the PBE+D3(BJ) level of theory.

| $T$ (K) | $a$ (Å) | $b$ (Å) | $c$ (Å) |
| --- | --- | --- | --- |
| 0 | 10.378 | 12.739 | 24.422 |
| 25 | 10.382 | 12.741 | 24.424 |
| 50 | 10.393 | 12.752 | 24.434 |
| 75 | 10.407 | 12.768 | 24.453 |
| 100 | 10.424 | 12.783 | 24.476 |
| 125 | 10.441 | 12.802 | 24.502 |
| 150 | 10.460 | 12.823 | 24.526 |
| 175 | 10.482 | 12.844 | 24.550 |
| 200 | 10.504 | 12.869 | 24.572 |
| 225 | 10.531 | 12.888 | 24.600 |
| 250 | 10.554 | 12.916 | 24.628 |
| 275 | 10.578 | 12.945 | 24.656 |
| 300 | 10.605 | 12.975 | 24.684 |
| 325 | 10.634 | 13.011 | 24.706 |
| 350 | 10.664 | 13.041 | 24.742 |

Comparison of the simplified QHA and the traditional QHA as implemented in PHONOPY

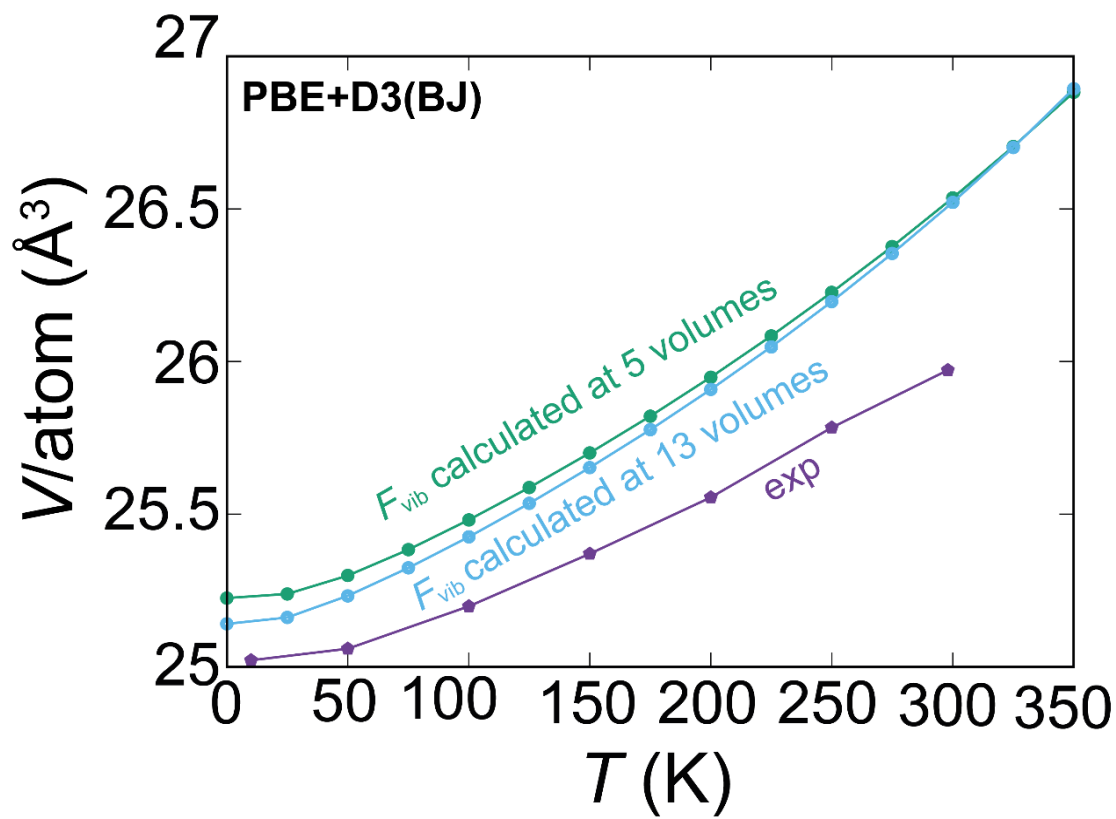

**Figure S4.** Thermal Expansion calculated with the quasi-harmonic approximation as implemented in PHONOPY[4,5] (blue) where we calculate $F_{vib}$ at 13 volumes (including both compressed and expanded volumes) and the quasi-harmonic approximation we used in the main text (green) at the PBE+D3(BJ) level. The differences between both approaches are smaller than 0.10 Å$^3$/atom. We used the Murnaghan equation of state[6] in both approximations.

ADPs in the quasi-harmonic approximation:

Equivalent displacement parameters ($U_{eq}$) of α-sulfur in the quasi-harmonic approximation

**Table S10.** Equivalent displacement parameters $U_{eq}$ (Å$^2$) calculated in the quasi-harmonic approximation.

| T(K) | vdW-DF2 | PBE+TS | PBE+D3(BJ) |
|---|---|---|---|
| 100 | 0.01985 | 0.01857 | 0.01590 |
| 150 | 0.03060 | 0.02770 | 0.02440 |
| 200 | 0.04261 | 0.03815 | 0.03412 |

Main-axis components ($U_1$, $U_2$, $U_3$) in the quasi-harmonic approximation

**Table S11.** Equivalent displacement parameters $U_{eq}$ (Å$^2$) calculated in the quasi-harmonic approximation.

|  | vdW-DF2 | | | PBE+TS | | | PBE+D3(BJ) | | |
|---|---|---|---|---|---|---|---|---|---|
| **100 K** | | | | | | | | | |
| S1 | 0.01206 | 0.01824 | 0.02888 | 0.01162 | 0.01780 | 0.02785 | 0.01019 | 0.01439 | 0.02213 |
| S2 | 0.01318 | 0.02299 | 0.02548 | 0.01337 | 0.01957 | 0.02535 | 0.01059 | 0.01795 | 0.02143 |
| S3 | 0.01438 | 0.02056 | 0.02423 | 0.01475 | 0.01955 | 0.01983 | 0.01140 | 0.01545 | 0.02049 |
| S4 | 0.01212 | 0.02048 | 0.02555 | 0.01124 | 0.01812 | 0.02383 | 0.01037 | 0.01496 | 0.02142 |
| **200 K** | | | | | | | | | |
| S1 | 0.02509 | 0.03821 | 0.06585 | 0.02304 | 0.03585 | 0.05944 | 0.02111 | 0.03001 | 0.04824 |
| S2 | 0.02800 | 0.05000 | 0.05618 | 0.02749 | 0.04011 | 0.05279 | 0.02207 | 0.03867 | 0.04741 |
| S3 | 0.03049 | 0.04421 | 0.05116 | 0.03011 | 0.03948 | 0.04133 | 0.02426 | 0.03273 | 0.04519 |
| S4 | 0.02539 | 0.04304 | 0.05372 | 0.02255 | 0.03751 | 0.04815 | 0.02115 | 0.03231 | 0.04631 |